\documentclass[a4paper, 11pt]{article}
\pdfoutput=1 
\usepackage{jheppub} 

\newcommand{\be}{\begin{equation}}
\newcommand{\ee}{\end{equation}}
\newcommand{\ba}{\begin{eqnarray}}
\newcommand{\ea}{\end{eqnarray}}

\newcommand{\la}{\label}

\newcommand{\eqn}[1]{(\ref{#1})}

\newcommand{\del}{\partial}

\title{\boldmath  The Gribov problem in Noncommutative QED}
\author[1]{Fabrizio Canfora} 
\author[2,3,5]{Maxim A.\ Kurkov} 
\author[4,5]{Luigi Rosa}
\author[4,5]{Patrizia Vitale}

\affiliation[1]{Centro de Estudios Cient\'{\i}ficos (CECS), Casilla 1469,
Valdivia, Chile.}
\affiliation[2]{Dipartimento di Matematica, Universit\`{a} di Napoli
{\sl Federico II}
\\
Monte S.~Angelo, Via Cintia, 80126 Napoli, Italy}
\affiliation[3]{
CMCC-Universidade Federal do ABC, Santo Andr\'e, S.P., Brazil}
\affiliation[4]{Dipartimento di Fisica, Universit\`{a} di Napoli
{\sl Federico II}\\
Monte S.~Angelo, Via Cintia, 80126 Napoli, Italy}
\affiliation[5]{INFN, Sezione di Napoli, \\
Monte S.~Angelo, Via Cintia, 80126 Napoli, Italy}

\emailAdd{ Canfora@cecs.cl} 
\emailAdd{kurkov@na.infn.it}
\emailAdd{ luigi.rosa@na.infn.it}
\emailAdd{patrizia.vitale@na.infn.it}

\abstract{
It is shown that  in the noncommutative version of QED {(NCQED)} Gribov copies induced by 
the noncommutativity
of space-time  appear in the Landau gauge. 
This is a genuine effect of noncommutative geometry which disappears when the noncommutative
parameter vanishes. 
}

\begin{document}
\maketitle
\flushbottom

\section{Introduction}
\label{sec:intro}
Space-time noncommutativity emerges at Plank scale in gedanken experiments when one tries to conjugate Quantum Mechanics and General Relativity \cite{DFR03}, \cite{Br12}. The result is the failure of the classical description of space-time as a pseudo-Riemannian manifold and the appearance of  uncertainty principles which are compatible with non-commuting coordinates. In different contexts, both string theory  \cite{stringsNC1}  and  loop quantum gravity 
\cite{As12}, \cite{Ro11}, 
predict the appearance of space-time noncommutativity at a fundamental level. 
It is therefore natural to investigate how gauge theories have to be modified in order to accomplish with the quantum  structure of space-time in extreme energy regimes.  

In gauge theory one of the fundamental problems to be solved, in order to be
able to perform perturbative computations of physical quantities, is the
overcounting of the degrees of freedom related to gauge invariance (see, for
instance, the detailed analysis in \cite{DeW03}). {{The
Faddeev-Popov gauge fixing }} procedure allows for perturbative computations
around the trivial vacuum $A_{\mu }=0$. On the other hand, the existence of
a proper gauge transformation preserving the gauge-fixing, would spoil the
whole quantization procedure. In \cite{Gri78} Gribov showed that in
non-Abelian gauge theories (on flat topologically trivial space-times) a 
\textit{proper gauge fixing} is not possible. Moreover, Singer \cite{singer}
showed that if Gribov ambiguities occur in the Coulomb gauge, they occur in all
the gauge fixing conditions\footnote{%
Other gauge fixings are possible such as the axial gauge, the temporal
gauge, etc., nevertheless these choices have their own problems (see, for
instance, \cite{DeW03}).} involving derivatives of the gauge field. The problem is generally addressed in the Landau gauge and in this paper we will stick to the latter, although the  Gribov-Zwanziger modification of the gauge action to cure the problem of Gribov copies has been recently extended from the Landau gauge to general $R_\xi$ gauges \cite{lechtenfeld}.

In the path integral formalism, a Gribov copy close to the identity of the
gauge group corresponds to a smooth zero mode of the Faddeev-Popov (\textbf{%
FP}) {{operator}}. In order to define the path integral in the
presence of Gribov copies close to the identity, the most successful method
is to restrict the path-integral to the neighborhood of $A_{\mu }=0$ in the
functional space of transverse gauge potentials, where the FP operator is
positive (see, in particular, \cite{Gri78} \cite{Zw82} \cite{DZ89}\ \cite%
{Zwa96} \cite{Va92}).When the space-time metric is flat,
this approach coincides with the usual perturbation theory and, at the same
time, it takes into account the infrared effects related to the (partial)
elimination of the Gribov copies \cite{Zw82} \cite{MaggS} \cite{Gracey}. If
one computes the propagator corresponding to such a restriction, one finds
the famous Gribov form factor for the propagator 
\begin{equation}
G^{\mathrm{G-Z}}(p)\sim \frac{p^{2}}{p^{4}+\gamma ^{4}},  \label{GZ}
\end{equation}%
where the dimensional constant $\gamma $ is related with the size of the
Gribov horizon. Although at high momenta such a propagator recovers the
usual one $\sim 1/p^{2}$, the infrared behavior is drastically different.
When one takes into account the presence of suitable condensates \cite{SoVar}
\cite{SoVar2} \cite{SoVar3} the agreement with lattice data is excellent 
\cite{DOV} \cite{soreprl}. This approach allowed to solve (see \cite{CaRo})
the well known sign problem of the Casimir energy and force in the MIT-bag
model.  Thus, in a sense, the
Gribov problem is not just a problem since, as the whole (refined)
Gribov-Zwanziger approach shows, it also suggests in a natural way a
solution which allows to go far beyond perturbation theory\footnote{%
In particular, both in the case of the glueballs mass spectrum and in the
case of (the solution of) the sign problem in the Casimir energy and force
in the MIT bag model, perturbation theory is obviously not sufficient to get
the correct answers as non-perturbative physics is needed.} in a very
successful way.

On the other hand, on a space-time with curved metric and/or non-trivial
topology the situation can be much more complicated since also Abelian gauge
theories can have smooth zero modes of the Faddeev-Popov operator \cite{CGO}, the maximally (super)symmetric vacuum can be \textit{outside} the Gribov
region \cite{ACGO} and the modular region could shrink to zero \cite{CGO2} 
\cite{giampi}.  For these reasons, it is natural to wonder whether the presence
of noncommutativity can induce Gribov copies even in U(1) gauge theories, which, because of noncommutativity of the product, develop self-interaction terms, thus behaving as non-Abelian gauge theories, with 
\be
F_{\mu\nu}= \del_\mu A_\nu -\del_\nu A_\mu -i [A_\mu, A_\nu]_\star. \label{field}
\ee
Noncommutativity thus modifies the covariant derivative (see Eq. \eqn{cd} below)  in the same way as non-Abelian gauge symmetry  and  it is then natural to expect a non-trivial equation for copies. 
This issue is extremely important since, as we already noticed,  noncommutative geometry at a
fundamental level has been shown to manifest in many different approaches to quantum gravity.  Consequently, it is mandatory to investigate whether noncommutative
geometry induces novel features in QFT which prevent  from using the
standard perturbative techniques.

The paper is organized as follows. In section 2 we establish the equation
for infinitesimal Gribov copies for noncommutative QED. In section 3 we
investigate the existence of exact solutions for particularly simple gauge
potentials and we actually show that we may have an infinite number of
genuine noncommutative solutions. In section 4 we discuss our results and
draw some conclusions.

\section{Equation for Gribov copies}

Let us fix the notations: for each two functions $f(x)$ and $g(x)$ the
noncommutative Moyal star product $f\star g$ is defined as follows 
\begin{equation}
(f\star g)(x)=f(x)\exp \left\{ \frac{i}{2}\,\theta ^{\rho \sigma }\overset{%
\leftarrow }{\partial _{\rho }}\overset{\rightarrow }{\partial _{\sigma }}%
\right\} g(x),  \label{star}
\end{equation}%
where the indices \footnote{%
Although it is not necessary, let us assume, that our space time is even
dimensional, since we are interested in $d=2$ and $d=4$.} $\rho ,\sigma =1,..,d$
and $d$ is the dimension of the space-time, which we assume Euclidean. The
antisymmetric matrix $\theta $, has the following nonzero components 
\begin{equation}
\theta _{1,2}=-\theta _{2,1}=\theta_1 ,\,\,\theta _{3,4}=-\theta _{4,3}=\theta_2
,\,\,...\,\,,\theta _{d-1,d}=-\theta _{d,d-1}=\theta_{d/2} ,  \label{nctheta}
\end{equation}%
where $\theta_i $ are real deformation parameters, in principle all different from each other, 
characterizing noncommutativity. A rescaling could be performed in order to make all parameters equal but we keep them different because it wouldn't simplify the calculations in the multidimensional case analyzed in  section \ref{multi}. When $\theta_i \rightarrow 0$, the star
product $\star $ goes to the standard commutative point-wise product of $f$
and $g$. 
\subsection{Gauge transformation}

\label{gauge} Under the $U(1)$ gauge transformation in NCQED the gauge field 
$A$ transforms as follows 
\begin{equation}
A \rightarrow A^{\prime}_{\mu}[\alpha] = U\star A_{\mu} \star U^{\dagger}
+i\, U\star \partial_{\mu} U^{\dagger}, \quad \quad U\equiv
\exp_{\star}\left(i \alpha\right),  \label{gtrans}
\end{equation}
where the star exponent of an arbitrary function $f$ is by definition 
\begin{equation}
\exp_{\star}(f) \equiv \sum_{n=0}^{\infty} \frac{1}{n!}\underbrace{%
f\star...\star f}_{\mbox{$n$ times}},
\end{equation}
and $\alpha$ is some function of $x$ considered as a parameter of the
transformation. It is worth noting that in the commutative limit $%
\theta\rightarrow 0$, the gauge transformation Eq.~\eqn{gtrans} reduces to
the standard Abelian gauge transformation 
\begin{equation}
A \rightarrow A^{\prime}_{\mu}[\alpha] = A_{\mu} + \partial_{\mu}\alpha + 
\mathcal{O}(\theta).  \label{gtranscomm}
\end{equation}
For us of crucial importance will be the infinitesimal form of the gauge
transformation Eq.~\eqref{gtrans}: 
\begin{equation}
A \rightarrow A^{\prime}_{\mu}[\alpha] = A_{\mu} + D_{\mu}\alpha + \mathcal{O%
}(\alpha),  \label{gtransinfin}
\end{equation}
where the covariant derivative $D_{\mu}$ appears due to \emph{non
commutativity} and is given by 
\begin{equation}
D_{\mu}f = \partial_{\mu}f + i \left(f\star A_{\mu} - A_{\mu}\star f\right)
\label{cd}
\end{equation}
for an arbitrary function $f(x)$.

\subsection{Zero mode equation in general}

Let us fix the gauge to be the Landau gauge, 
\begin{equation}
\partial^{\mu}A_{\mu} = 0.  \label{gf}
\end{equation}
In commutative QED this gauge fixing condition fixes the gauge completely, indeed, under suitable regularity conditions at the boundary,
 if $A_{\mu}$ satisfies Eq.~\eqref{gf}, the transformed field under Eq.~%
\eqref{gtranscomm} automatically does not 
\begin{equation}
\partial^{\mu}A_{\mu}^{\prime}[\alpha] \neq 0, \quad \alpha\neq 0.
\end{equation}
In other words the equation 
\begin{equation}
\partial^{\mu}A_{\mu}^{\prime}[\alpha] = 0  \label{zmeq}
\end{equation}
called also the ``zero mode equation" has only the trivial solution $\alpha=0
$. However, in a more general setting, for example non-Abelian gauge
theories, the zero mode equation Eq.~(\ref{zmeq}) may have nontrivial
solutions $\alpha \neq 0$ called \emph{zero modes}.

The lack or the presence of zero modes means correspondingly the possibility
or impossibility to foliate the functional space of all possible gauge
fields into the orbits of the gauge group {{in a way, that each
gauge orbit intersects the gauge fixing hypersurface Eq.~\eqref{gf} just
once}} ({{in order to be able to}} integrate over the
representatives of each equivalence class). The latter situation leads to an
overcounting of degrees of freedom when one performs the functional integral
over $A$, creating the Gribov problem in the infrared regime.

The main goal of the present research is to figure out whether the zero mode
equation Eq.~\eqref{zmeq} can exhibit nontrivial solutions in the case of
NCQED, i.e. when $A^{\prime}[\alpha]$ is given by Eq.~\eqref{gtrans}.

\subsection{Infinitesimal Gribov copies}

The \emph{infinitesimal zero mode equation} which corresponds to the
infinitesimal gauge transformations Eq.~\eqref{gtransinfin} is of special
interest since it has direct relation with  the Faddeev-Popov ghost action and
with  the Gribov-Zwanziger term. 

Substituting Eq.~\eqref{gtransinfin} in the general formula Eq.~\eqref{zmeq}
we obtain 
\begin{equation}
\partial^{\mu}D_{\mu}\alpha = 0.  \label{zminfin}
\end{equation}
Let us understand the structure of this equation from the mathematical point
of view. Substituting the expression of the covariant derivative Eq.~%
\eqref{cd} and the star product Eq.~\eqref{star} into Eq.~\eqref{zminfin} we
arrive at the following zero mode equation written in terms of $\alpha$ and
its derivatives 
\begin{equation}
-\partial^2 \alpha + \underbrace{iA_{\mu} \exp\left\{ \frac{i}{2}%
\,\theta^{\rho\sigma}\overset{\leftarrow}{\partial_{\rho}}\overset{%
\rightarrow}{\partial_{\sigma}}\right\} (\partial^{\mu}\alpha) -
i(\partial^{\mu}\alpha)\exp\left\{ \frac{i}{2}\,\theta^{\rho\sigma}\overset{%
\leftarrow}{\partial_{\rho}}\overset{\rightarrow}{\partial_{\sigma}}\right\}
A_{\mu}}_{\mbox{nonlocal terms}} = 0.  \label{zminfinfull}
\end{equation}
The presence of nonlocal terms implies that, differently form QCD, this is
not a differential equation and its resolution is a very hard task. However,
in order to say whether we have Gribov copies or not we only need to
understand whether it has nontrivial solutions $\alpha\neq0$.

On performing the  Fourier transform
of $\alpha$ and $A$  one can rewrite the pseudo-differential equation Eq.~%
\eqref{zminfinfull} as a homogenous Fredholm equation of the second kind. After 
some simple computations we obtain indeed
\begin{eqnarray}
0 &=&-\partial ^{2}\alpha +iA_{\mu }\exp \left\{ \frac{i}{2}\,\theta ^{\rho
\sigma }\overset{\leftarrow }{\partial _{\rho }}\overset{\rightarrow }{%
\partial _{\sigma }}\right\} (\partial ^{\mu }\alpha )-i(\partial ^{\mu
}\alpha )\exp \left\{ \frac{i}{2}\,\theta ^{\rho \sigma }\overset{\leftarrow 
}{\partial _{\rho }}\overset{\rightarrow }{\partial _{\sigma }}\right\}
A_{\mu }  \notag \\
&=&\int d^{d}ke^{ikx}\left\{ -k^{2}\hat{\alpha}(k)+2i\,\int d^{d}q\sin
\left( -\frac{1}{2}\theta ^{\rho \sigma }q_{\rho }k_{\sigma }\right) k^{\mu }%
\hat{A}_{\mu }(q)\hat{\alpha}(k-q)\right\} 
\end{eqnarray}%
which is equivalent to the following integral equation 
\begin{equation}
k^{2}\hat{\alpha}(k)+2i\,\int d^{d}q\sin \left( \frac{1}{2}\theta ^{\rho
\sigma }q_{\rho }k_{\sigma }\right) k^{\mu }\hat{A}_{\mu }(q)\hat{\alpha}%
(k-q)=0.  \label{ieq}
\end{equation}%
Changing the integration variable $q\rightarrow k-q$ we finally arrive at 
\begin{equation}
\hat{\alpha}(k)=\int d^{d}q\,\,Q(q,k)\,\,\hat{\alpha}(q),  \label{ieq2}
\end{equation}%
which is a  homogeneous Fredholm equation of the second kind, with the 
 kernel $Q$ given by 
\begin{equation}
Q(q,k)=-\frac{2i\,k^{\mu }\hat{A}_{\mu }(k-q)}{k^{2}}\sin \left( \frac{1}{2}%
\,\theta ^{\rho \sigma }q_{\rho }k_{\sigma }\right) .  \label{Qdef}
\end{equation}%
\footnote{\label{Coulombgauge} Let us notice that, in the Coulomb gauge  $\del_j A^j= 0, j=1,...,d-1$ 
the equation for the copies is formally the same as  eq. \eqn{ieq}, provided we replace $k^2$ with $\vec k^2$ and $k^{\mu }\hat{A}_{\mu }(q)$ with $k^{j }\hat{A}_{j }(q)$. 
Obviously the solutions will be different. In particular none of the Ans\"atze we shall make in the rest of the paper can be easily adapted to the Coulomb gauge. We shall 
comment more on this aspect in the discussion session.}
It is possible to recast the integral equation Eq.~\eqref{ieq2} in such a
way that the corresponding integral operator becomes manifestly symmetric.
To this purpose we notice that, due to the gauge fixing condition, $\hat{A}%
_{\mu }(k)k^{\mu }=0$, one can replace $k_{\mu }$ by $(k_{\mu }+q_{\mu })/2$
in Eq.~\eqref{Qdef}. Upon making the change of variable $\beta =|k|\cdot
\alpha $, with $|k|\equiv \sqrt{k_{\mu }k^{\mu }}$ and multiplying both
sides of Eq.~\eqref{ieq2} by $k$, we arrive at the following equivalent
integral equation: 
\begin{eqnarray}
\hat{\beta}(k) &=&\int d^{d}q\,\,P(q,k)\,\,\hat{\beta}(q), \\
P(q,k) &=&-\frac{i\,\left( k^{\mu }+q^{\mu }\right) \hat{A}_{\mu }(k-q)}{%
|k||q|}\sin \left( \frac{1}{2}\,\theta ^{\rho \sigma }q_{\rho }k_{\sigma
}\right) .  \label{ieq3}
\end{eqnarray}%
In order to show that the linear integral operator $P$ defined by the kernel
Eq.~\eqref{ieq3} is formally self-adjoint it is necessary and sufficient to
show that the corresponding kernel satisfies 
\begin{equation}
P^{\ast }(q,k)=P(k,q),  \label{hermiticity}
\end{equation}%
where \textquotedblleft *" means complex conjugation. Now we recall that we
are interested in \emph{real} gauge potentials $A(x)$, which is equivalent
to impose 
\begin{equation}
\hat{A}^{\ast }(k)=\hat{A}(-k)  \label{Aconj}
\end{equation}%
on the corresponding Fourier transform. Performing complex conjugation of
Eq.~\eqref{ieq3} and using Eq.~\eqref{Aconj} and skew symmetry of $\theta $
we immediately obtain 
\begin{equation}
\left( \frac{i\,\left( k^{\mu }+q^{\mu }\right) \hat{A}_{\mu }(k-q)}{|k||q|}%
\sin \left( \frac{\theta ^{\rho \sigma }}{2}\,q_{\rho }k_{\sigma }\right)
\right) ^{\ast }=\frac{i\,\left( q^{\mu }+k^{\mu }\right) \hat{A}_{\mu }(q-k)%
}{|q||k|}\sin \left( \frac{\theta ^{\rho \sigma }}{2}\,k_{\rho }q_{\sigma
}\right) 
\end{equation}%
that is exactly the equality Eq.~\eqref{hermiticity}.

In principle self-adjoint operators have an infinite set of eigenfunctions
and eigenvalues, however since we are in the infinite dimensional situation
a lot depends on the properties of the kernel Eq.~\eqref{ieq3}. 
If for some particular $\hat A(k) = B(k)$, a complete set of
eigenfunctions $\psi_n$ with eigenvalues $\lambda_n$ exists,  we
obtain
\begin{equation}
\psi_n (k) = \frac{1}{\lambda_n} \int d^d q \,P(q,k)| _{A=B}\psi_n(q), \quad
n = 1,2...
\end{equation}
The latter implies that we have an infinite set of gauge potentials $%
B/\lambda_n,~ n=1,2...$ and each of them exhibits zero modes $\alpha = \psi_n
$.

\subsection{The Henyey approach}

As it is done in standard QCD, also in this case one can follow the Henyey
strategy \cite{heynyey} where one fixes the form of the zero modes $\hat{%
\alpha}(k)$ and solves for the gauge potential.  

It is worth emphasizing here that, in
the standard commutative case, the Henyey strategy\ to fixing the form of the
copies and to solving for the gauge potential gives rise to algebraic equations
which can be easily solved while, in the present noncommutative case, even
following such a strategy leads to a rather non-trivial equation, due to the
non-locality appearing in noncommutative geometry. Therefore we will only sketch the strategy here but we will follow a different approach in the next section. 

As we will show, in the noncommutative Henyey
approach, instead of a homogeneous Fredholm equation of the second kind we obtain  a non-homogeneous Fredholm equation of the first kind, with
 Hermitian integral kernel. 

Since $\hat A_{\mu}$ has $d$ components constrained by the gauge fixing
condition, but the integral equation is an equation for  one
unknown function, we are free to choose some particular parametrization of $%
\hat A$. Let us make the following  Ansatz: 
\be
\hat A_{\mu}(q) = i \,a(q)G_{\mu}(q) ~~~ ~\mbox{with} \, G_\mu(q) =
-\tilde \theta_{\mu\nu} q^\nu 
 \label{ans}
\ee
with $\tilde \theta$ the inverse matrix of the matrix $\theta$. This   obviously satisfies the gauge condition  $q^{\mu}\hat A_{\mu}(q) =0$. The presence of the
imaginary unit in the first line of Eq.~\eqref{ans} will become clear soon.

Substituting the Ansatz Eq.~\eqref{ans} in Eq.~\eqref{ieq} we get the
following (non homogeneous) Fredholm equation of the first kind for the unknown function $a(q)$: 
\begin{equation}
\int d^d q \,R(q,k) a(q) = f(k), \label{aeq}
\end{equation}
where 
\begin{eqnarray}
&& f(k) = k^2\hat\alpha(k), \quad\mbox{and}  \notag \\
&& R(q,k) = 2 \sin\left(\frac{1}{2}\,\theta^{\rho\sigma}k_{\rho}q_{\sigma}%
\right) k^{\mu}G_{\mu}\,\, \hat\alpha(k-q).  \label{ieq5}
\end{eqnarray}
Note that 
\begin{equation}
k^{\mu}G_{\mu}(q) = -k^\mu \tilde\theta_{\mu\nu} q^\nu 
\end{equation}
 is \emph{real} and skew symmetric with respect to the
exchange $k\leftrightarrow q$: 
\begin{equation}
k^{\mu}G_{\mu}(q) = -q^{\mu}G_{\mu}(k)   \label{kqprop}
\end{equation}
therefore the combination 
\begin{equation*}
\sin\left(\frac{1}{2}\,\theta^{\rho\sigma}k_{\rho}q_{\sigma}\right)
k^{\mu}G_{\mu} 
\end{equation*}
is real and symmetric with respect to the mentioned exchange. Since we
are interested in \emph{real} $\alpha(x)$, the corresponding Fourier
transform $\hat\alpha(k)$ satisfies
$ 
\left(\hat\alpha(k)\right)^* = \hat\alpha(-k)$. Summarizing all observations listed above, we conclude that the kernel $%
R(k,q)$defined by Eq.~\eqref{ieq5} satisfies 
\begin{equation}
\left(R(k,q)\right)^* = R(q,k),
\end{equation}
therefore the corresponding linear integral operator $R$ is \emph{self-adjoint}. Exactly for this reason the imaginary unit in the Ansatz Eq.~%
\eqref{ans} is needed.

In conclusion, in order to solve Eq. \eqn{aeq} for the potential it is sufficient to show that for
some particular choice of $\alpha$ the inverse integral operator $R^{-1}$
exists, so that $a = R^{-1} f$ will give us the gauge potential, for which $%
\alpha$ is a zero mode.

\section{Some exact solutions.}

The question of the existence of Gribov copies is equivalent to the question of
the existence of eigenvectors of the self-adjoint operator defined by Eq.~%
\eqref{ieq3}. No matter how we choose the gauge potential $A_{\mu}$,  this
operator \emph{does not} belong to the Hilbert-Schmidt class, since  
\begin{equation}
\int \,dq \,dk\, |P(q,k)|^2 = \infty,
\end{equation}
so one can not say a priori whether there exists at least one gauge
potential exhibiting zero modes! Neither one can say the opposite. The aim
of this section is to demonstrate that gauge potentials that give solutions
of Eq.~\eqref{ieq3} \emph{do} exist.

For the scope of the present section we will not use the property of
Hermiticity of $P$, therefore it will be more convenient for the forthcoming
computations to resort to the original form of the zero mode equation Eq.~%
\eqref{ieq}, which we rewrite as 
\begin{equation}
k^2 \hat \alpha(k) +2 i k^{\mu} \int d^d q \,\sin{\left(\frac{1}{2}%
\theta^{\sigma\rho}q_{\rho}k_{\sigma}\right)}\hat
A_{\mu}(k-q)\,\hat\alpha(q)=0.  \label{iEQ}
\end{equation}
We notice that if we consider gauge potentials $\hat A_{\mu}$ which are
proportional to derivatives of $\delta(k)$ , Eq. (\ref{iEQ}) becomes a
differential equation for $\hat\alpha(k)$.

\subsection{The simplest situation}

First we try the following Ansatz 
\begin{equation}
A_{\mu} = Q \tilde\theta_{\mu\nu} x^{\nu}  \label{Acoord}
\end{equation}
with $Q$ some constant to be fixed. 
 The Fourier transform reads 
\begin{equation}
\hat A_{\mu}(k) = i Q \tilde\theta_{\mu\nu} \partial^{\nu}\delta(k).
\label{p1}
\end{equation}

This potential obviously satisfies the gauge fixing condition $%
\partial^{\mu}A_{\mu}$ in coordinate space, while in momentum space we deal
with a distribution, therefore we have to specify in which sense the
equality 
\begin{equation}
k^{\mu}\hat A_{\mu}(k) = 0  \label{gfcm}
\end{equation}
holds. Let us fix the space of probe functions $\hat\alpha(k)$ to be the
Schwartz space of infinitely smooth functions decaying at infinity faster
than any arbitrary power. For an arbitrary Schwartzian function $%
\hat\alpha(k)$ one must have 
\begin{equation}
\int d^d k\, \hat\alpha(k) \,k^{\mu}\hat A_{\mu} = 0  \label{gfm}
\end{equation}
which is satisfied by Eq. (\ref{p1}) it being for arbitrary $\hat \alpha(k)$ 
\begin{eqnarray}
&& \int d^d k\, \hat\alpha(k) \,k^{\mu}i Q\tilde\theta_{\mu\nu}
\partial^{\nu}\delta(k) \equiv - i Q \tilde\theta_{\mu\nu}\left[%
\partial^{\nu}(k^{\mu}\hat\alpha(k))\right]\big |_{k=0}  \notag \\
&& =-iQ(\underbrace{\tilde\theta_{\mu\nu}\delta^{\mu\nu}}_0 \hat\alpha(0)
+\tilde\theta_{\mu\nu}(\partial^{\nu}f)\big |_{k=0} \cdot\underbrace{k^{\mu}%
\big |_{k = 0}}_0 )= 0.
\end{eqnarray}
The reason to search for zero modes $\alpha(x)$ belonging to the Schwarz
space is twofold. On one side we observe that, in the commutative case, the
zero mode equation is a Laplace equation which, unless one specifies
boundary conditions, may have nontrivial solutions. Indeed each linear
function solves it. On the other side the Green function of the
corresponding Laplacian gives a singular solution, that decreases at
infinity for $d > 2$. In order to get rid of these irrelevant solutions (in
the commutative case there is no Gribov problem!) we impose regularity of $%
\alpha$ at each finite point and vanishing at infinity, which are both satisfied by  Schwarz functions. In this class of functions the commutative zero mode equation
has just the trivial solution $\alpha = 0$.

Substituting the ansatz \eqref{p1} in the equation \eqref{iEQ} and using 
\begin{equation}
-2Qk^{\mu}\tilde\theta_{\mu\nu} \int d^d q \,\sin{\left(\frac{1}{2}%
\theta^{\sigma\rho}q_{\rho}k_{\sigma}\right)} \hat \alpha(q)\,
^{q}\partial^{\nu} \delta(k-q) = Q k^2\hat \alpha(k),
\end{equation}
we arrive at the following algebraic equation 
\begin{equation}
(1+ Q)k^2 \hat\alpha(k) = 0,
\end{equation}
which exhibits nontrivial solutions. Indeed if (and only if) 
\begin{equation}
Q = -1 ,
\end{equation}
for arbitrary even space-time dimension, any \emph{arbitrary} function $\hat\alpha(k)$ is a solution! Unfortunately,
although we found nontrivial solutions of Eq.~\eqref{iEQ}, this particular
gauge potential has a peculiar feature. One may show \cite{wallet} that it
is invariant under gauge transformations \eqref{gtrans} and therefore we
do not have Gribov copies.

Nevertheless this potential is of interest. First of all,  the existence of such a gauge invariant connection is a purely
noncommutative feature \cite{wallet} (also see \cite{MVW13} where such a connection has been used to study NCQED as a nonlocal matrix model) and does not exist in the commutative
limit. Second, its smooth approximations may be used in principle to search
solutions of the integral equation Eq.~\eqref{iEQ}.

\subsection{The next to the simplest situation}

To simplify the presentation let us consider the two dimensional case.  Here we have only one noncommutative parameter, $\theta_{12}=-\theta_{21}=\theta$. The
next to the simplest gauge potential leading to a viable differential
equation is the following one:\footnote{ In principle one may
also consider quadratic potentials, however for technical reasons we prefer
to deal with rotationally invariant potentials, therefore the next to the
simplest potential which we consider is cubic. Indeed, in $d=2$ rotationally
invariant quadratic in $x$ gauge fields do not exist.}
\begin{equation}
A_{\mu}(x) \propto \tilde\theta_{\mu\nu} x^{\nu} x^2,  \label{ntsp}
\end{equation}
which, being in two dimensions, can be further simplified to the form
\be
A_{\mu}(x)= Q \varepsilon_{\mu\nu} x^{\nu} x^2,  \label{ntsp2}
\ee
with $Q$ some constant to be determined and $\varepsilon_{\mu\nu}$ the Levi-Civita tensor in two dimensions.  The 
corresponding Fourier transform reads 
\begin{equation}
\hat A_{\mu} (k) = i Q \varepsilon_{\mu\nu}\,\square\, \partial^{\nu}
\delta(k).  \label{p2}
\end{equation}
It is worth emphasizing here that the gauge potential in Eq. (\ref{p2}) can
be approximated as closely as one wants replacing the $\delta $-function
with a Gaussian (obviously, if one would use the Gaussian from the very
beginning the Gribov copies equation would not be solvable anymore). Hence,
the present example is not only interesting in itself, since it also shows
that there is a whole family of smooth gauge potentials which are
arbitrarily close to having smooth normalizable Gribov copies.

In what follows we will refer to $Q$ as the amplitude of the potential. In
spatial  coordinates it obviously satisfies the Landau gauge fixing
condition and for consistency we check whether it satisfies the gauge fixing
condition \eqref{gfcm} in the above mentioned ``distributive" sense. For an
arbitrary probe function $\hat\alpha(k)$ one obtains 
\begin{eqnarray}
&&\int d^d k\, \hat \alpha(k) k^{\mu}\hat A_{\mu} = i Q \varepsilon_{\mu\nu}
\int d^dk\, \hat\alpha \, k^{\mu} \square \,\partial^{\nu} \delta(k) =
-iQ\varepsilon_{\mu\nu}\left[\square\, \partial^{\nu}\left( \hat \alpha
k^{\mu}\right)\right]\big|_{k=0}  \notag \\
&& = -iQ ( \varepsilon_{\mu\nu}\left(\square\partial^{\nu}\hat\alpha\right)%
\big |_{k=0} \underbrace{k^{\mu} \big |_{k=0}}_{0} + 2(\underbrace{%
\varepsilon_{\mu\nu}\partial^{\mu}\partial^{\nu}}_0\hat\alpha )\big |_{k=0}
+ (\square\hat\alpha)\big |_{k=0}\underbrace{\varepsilon_{\mu\nu}\delta^{\mu%
\nu}}_0 ) =0.  \notag
\end{eqnarray}
Let us now substitute the potential Eq.~\eqref{p2} in the integral equation
Eq.~\eqref{iEQ} in order to derive a partial differential equation for the
zero modes $\hat \alpha(k)$. We obtain 
\begin{eqnarray}
&& Q k^{\mu}\epsilon_{\mu\nu} \int d^d q \left( ^{q}\square\,
^{q}\partial^{\nu}\, \delta(q-k) \right) \,\sin{\left(\frac{1}{2}%
\theta^{\sigma\rho}q_{\rho}k_{\sigma}\right)}\, \hat \alpha(q)  \notag \\
&&\; = - Q k^{\mu} \varepsilon_{\mu\nu} \left\{ ^{q}\square\,
^{q}\partial^{\nu} \left[ \sin{\left(\frac{1}{2}\theta^{\sigma\rho}q_{%
\rho}k_{\sigma}\right)}\, \hat \alpha(q) \right] \bigg |_{q=k} \right\} 
\notag \\
&&\;= \frac{Q\theta}{8}\left( \theta^2 k^4 \hat\alpha - 4 k^2 \square
\hat\alpha - 8\,\varepsilon^{\mu\nu}\varepsilon^{\eta\lambda}k_{\mu}
k_{\eta}\partial_{\nu}\partial_{\lambda}\hat\alpha\right)
\end{eqnarray}
hence the zero modes $\hat\alpha(k)$have to satisfy the partial differential
equation given below: 
\begin{equation}
\left(-4k^2\square - 8\,
\varepsilon^{\mu\nu}\varepsilon^{\eta\lambda}k_{\mu}k_{\eta}\partial_{\nu}%
\partial_{\lambda} - \frac{4k^2}{Q\theta} + \theta^2 k^4\right)\hat\alpha(k)
= 0.  \label{PDE}
\end{equation}
We notice that, since in two dimensions $\varepsilon^{\mu\nu}$ is a
universal tensor, this equation is rotationally invariant, therefore it
makes sense to rewrite it in polar coordinates $(r,\phi)$ given by 
\begin{equation}
\begin{cases}
k_1 = r \cos{\phi} \\ 
k_2 = r \sin{\phi}%
\end{cases}%
.
\end{equation}
One may easily see that in polar coordinates Eq.~\eqref{PDE} reads 
\begin{equation}
r^2 \hat\alpha_{rr} + 3 r\hat\alpha_r + \frac{1}{Q\theta}r^2\hat\alpha - 
\frac{\theta^2}{4}{r^4}\hat\alpha + 3\hat\alpha_{\phi\phi} = 0,
\end{equation}
therefore it exhibits separation of variables. Let us look for a solution in
the following form 
\begin{equation}
\alpha(\phi,r) = \Phi(\phi)\,f(r),
\end{equation}
where the functions $\Phi$ and $f$ satisfy the following \emph{ordinary}
differential equations 
\begin{equation}
\begin{cases}
-\Phi_{\phi\phi} = \lambda\Phi, \\ 
r^2 f_{rr} + 3r f_r + \left(-3\lambda + \frac{1}{Q\theta}r^2 - \frac{\theta^2%
}{4}r^4\right)f = 0.%
\end{cases}
\label{odes}
\end{equation}
The former is just the equation of the simple harmonic motion, while the
latter is the confluent hypergeometric equation whose properties are very
well studied, see e.g. \cite{Bat} for a review. Let us specify the boundary
conditions as follows: 
\begin{equation}  \label{bcs}
\begin{cases}
\Phi(0) = \Phi(2\pi), \\ 
|f(0)| <\infty, \quad f(r)\rightarrow 0, \, \mbox{when}\, r\rightarrow\infty.%
\end{cases}%
\end{equation}
Below we will see that each function satisfying the boundary conditions
given by Eq.~\eqref{bcs} belongs to the Schwarz space. We also notice that
the deformation parameter $\theta$ enters the equation Eq.~\eqref{odes} in
the combinations $Q\theta$ and $\theta^2$, where $Q$ is arbitrary. Therefore
without loss of generality one may consider only $\theta>0$, since the
opposite sign of $\theta$ corresponds to the opposite sign of the arbitrary
amplitude $Q$. From the angular boundary conditions we see that 
\begin{equation}
\lambda_n = n^2, n= 0, \pm1, \pm 2,...  \label{lambd}
\end{equation}
so that the general solution for the angular equation Eq.~\eqref{odes} is of
the form 
\begin{equation}
\Phi(\phi) = \tilde c_1 \cos{(n\phi)} + \tilde c_2 \sin{(n\phi)},
\end{equation}
where $\tilde c_1$ and $\tilde c_2 $ are arbitrary constants. The general
solution of the radial equation for $\lambda = \lambda_n$ is given by 
\begin{eqnarray}
&& f(r) = r^{\sqrt{3n^2 + 1} - 1} \exp{\left(-\frac{r^2 \theta}{4}\right)}%
\left(c_1\, M\left(a, c, \frac{\theta \,r^2}{ 2}\right) + c_2\,U\left(a, c, 
\frac{\theta \,r^2}{ 2}\right)\right),  \notag \\
&&\mbox{with}\quad a = \frac{1}{2} + \frac{1}{2}\sqrt{3n^2+1}-\frac{1}{%
2\theta^2 Q}, \quad \quad c = 1+\sqrt{3n^2 + 1}  \label{rsol}
\end{eqnarray}
$U$ and $M$ are Kummer functions and $c_1$ and $c_2$ are arbitrary constants.

We notice however that the boundary condition for the radial dependence can be
satisfied if and only if the number $a$ defined by Eq.~\eqref{rsol} is a non
positive integer 
\begin{equation}
a=-m,\quad m=0,1,2,...  \label{arestrict}
\end{equation}%
In that case Kummer functions reduce to Laguerre polynomials.
Thus, the solution regular both at zero and infinity (see \cite{Bat}) is %
\begin{equation}
f(r) = C\, r^{\sqrt{3n^2+1}-1}\exp{\left(-\frac{r^2 \theta}{4}\right)}\,
L_m^{\sqrt{3n^2 + 1}}\left(\frac{\theta\, r^2}{2}\right),
\end{equation}
where $C$ is an arbitrary constant, {{and $L_n^a(z)$ stands for
generalized Laguerre polynomial}}. This solution exists when the amplitude Q
takes one of the discrete values 
\begin{equation}
Q_{nm} = \frac{1}{\theta^2(\sqrt{3n^2 +1} + 2m+1)}, \quad n = 0, \pm1,
\pm2,...,\quad m = 0, 1, 2, ...  \label{Qnm}
\end{equation}

The general form of the zero modes, when the amplitude $Q$ belongs to the
discrete set defined above, is 
\begin{eqnarray}
&&\!\!\!\!\!\!\!\!\!\!\!\!\ \hat\alpha_{nm}(r,\phi) = \left(C_1\cos{(n\phi)}
+ C_2\sin{(n\phi)} \right) r^{\sqrt{3n^2+1}-1}\exp{\left(-\frac{r^2 \theta}{4%
}\right)}\, L_m^{\sqrt{3n^2 + 1}}\left(\frac{\theta\, r^2}{2}\right)  \notag
\\
&&\mbox {where}\quad C_1,C_2 \,\, \mbox{are real if} \,\, n\,\,\mbox{is even}
\notag \\
&& \quad \mbox{and} \quad C_1,C_2 \,\, \mbox{are purely imaginary if} \,\,
n\,\,\mbox{is odd}.  \label{SOLUT}
\end{eqnarray}
The restriction on arbitrary constants comes from the requirement to have
real $\alpha(x)$. Indeed, in order to satisfy this requirement, the
corresponding Fourier transform $\hat\alpha$ must satisfy $(\hat\alpha(k))^* = \hat\alpha(-k)$. From another side the reflection $k\rightarrow -k$ is
equivalent to the shift $\phi \rightarrow \phi +\pi$ in polar coordinates.
One may easily check that the radial dependence is real, and the linear
combination of sine and cosine appearing in \eqref{SOLUT} satisfies 
\begin{equation}
\left(C_1\cos{(n(\phi +\pi))} + C_2\sin{(n(\phi +\pi))} \right) =
\left(C_1\cos{(n\phi)} + C_2\sin{(n\phi)} \right)^*
\end{equation}
if and only if the restriction described in \eqref{SOLUT} is imposed.

\subsubsection{How many zero modes do we have?}

\label{howmany} In this subsection we discuss how many linearly independent
solutions for a given amplitude \eqref{Qnm} we have, connecting our results
with number theory. From the solution \eqref{SOLUT} we can guarantee at
least two. It is remarkable that for some special subset of the amplitudes $%
Q_{nm}$ we can have more than two, or more precisely the following statement
holds: \, \newline
\vskip.1cm \noindent\textbf{Theorem.} \textit{For arbitrarily large number N
there exists such an amplitude $Q_{nm}$, that the number of linearly
independent zero modes corresponding to this potential is greater than $N$.}
\, \newline
\vskip.1cm \noindent Before we go ahead to prove the existence of ``many"
solutions, let us remind some facts from number theory. The key property
allowing such an existence is due to the fact that there exist infinitely
many natural numbers $p_l, \, l=1,2,...$ such that 
\begin{equation}
{3p_l^2 + 1} \quad \mbox{is a perfect square},
\end{equation}
i.e. there exists such a natural number $P_l$, that 
\begin{equation}
{3p_l^2 + 1} = P_l^2, \quad l=1,2,...
\end{equation}
This sequence, called ``A001353", is well known and studied (see \cite%
{A001353} and refs therein). We have 
\begin{equation}
p_1 = 0,\, p_2 = 1,\, p_l = 4p_{l-1} - p_{l-2}, \quad l=3,4,...
\end{equation}
and the corresponding $P_l$ are given by 
\begin{equation}
P_l =2p_l - p_{l-1}.
\end{equation}
We also notice that all $P_{2k}, k=1,2,...$ are \emph{even} natural
integers, therefore 
\begin{equation}
J_k \equiv \frac{P_{2k}}{2}, \quad k=1,2,...
\end{equation}
are natural integers. 
\begin{table}
\centering
    \begin{tabular}{  c| c|c  }
   $l$ & $p_l$ & $P_l$ \\ \hline
   1   & 0        & 1 \\ \hline
   2 & 1 & 2 \\ \hline
   3 & 4 & 7 \\ \hline
   4 & 15 & 26 \\ \hline
   5 & 56 & 97 \\ \hline
   6 & 209 & 362 \\ \hline
   ... & ...& ...
   \end{tabular}
\caption{\label{T1} The table shows the first six numbers from the A001353 sequence. }
\end{table}
The first six numbers from the sequence A001353 are presented in
Table ~\ref{T1}.

We are now ready to prove the above stated theorem. For a given (arbitrary
large) N, we choose the amplitude $Q =Q_N$ where 
\begin{equation}
Q_N = \frac{1}{\theta^2\left(2J_N +1\right)}.  \label{QN}
\end{equation}
This implies that the parameter $a$ appearing in Eq.~\eqref{rsol} 
\begin{equation}
a = \frac{1}{2} + \frac{\sqrt{3n^2 + 1}}{2} - \frac{1}{2\theta^2 Q} = -J_N + 
\frac{\sqrt{3n^2+1}}{2}
\end{equation}
is a nonpositive integer if and only if $\frac{\sqrt{3n^2+1}}{2}$ is a
positive integer not greater than $J_N$. Let us now substitute $n=n_k$ given
by 
\begin{equation}
n_k = p_{2k},\quad k=1,..,N.
\end{equation}
We obtain 
\begin{equation}
a = -J_N + J_k,\quad k=1,...,N,
\end{equation}
and for each $k$ from 1 to N we have two linearly independent solutions
(with sine and cosine), given by \eqref{SOLUT}, where the Laguerre
polynomial is labelled by $m = -J_N + J_k$, and the parameter $\sqrt{3n^2+1} =
P_k$. Finally we conclude that we have found the amplitude $Q = Q_N$ for the
gauge potential in (\ref{ntsp2}), that exhibits $2N$ linearly independent
zero modes. {\it QED}

The logic of our construction is summarized in Table~\ref{T2}.

\begin{table}
\begin{center}
    \begin{tabular}{  l | l | l | p{0.5cm} }
 
    $k$ & $m$ & $\sqrt{3n^2+1}$ & n \\ \hline
    1 & $J_N - 1$ & 2 &  1 \\ \hline
    2 & $J_N - 13$ & 26 &  15 \\ \hline
    3 & $J_N - 181$ & 362 &  209 \\ \hline
    ... & ... & ... &  ... \\ \hline
   k & $J_N - J_K$ & $P_K$ & $p_{2k}$  \\ \hline
    ... & ... & ... &  ... \\ \hline
   N & $ 0$ & $P_N$ &  $p_{2N}$
    \\

    \end{tabular}

\end{center}
\caption{This table summarizes the construction of $N$ different solutions for a given amplitude, as described  in  section \ref{howmany}. \la{T2}}
\end{table}

\subsubsection{Examples}

Let us illustrate the results of the previous section with some examples.%
\newline

\begin{itemize}
\item {\ N=1.} The corresponding amplitude $Q$ is given by (see Eq.~%
\eqref{QN})  
\begin{equation}
Q_1 = \frac{1}{3\theta^2}.
\end{equation}
There are just two linearly independent solutions \newline
\begin{equation*}
m=0, \quad n=1 \quad \mbox{angular dependence = cosine}
\end{equation*}
\begin{equation*}
m=0, \quad n=1 \quad \mbox{angular dependence = sine}\;\;\,
\end{equation*}
In Fig. \ref{Fig1} we plot the corresponding potential for the cosine case%
\footnote{%
Since for odd $n$ the Fourier image $\hat\alpha(k)$ is purely imaginary, we
multiplied it by $i$ to build a plot.}. 
\begin{figure}[htb]
\centering
\includegraphics[scale = 0.33]{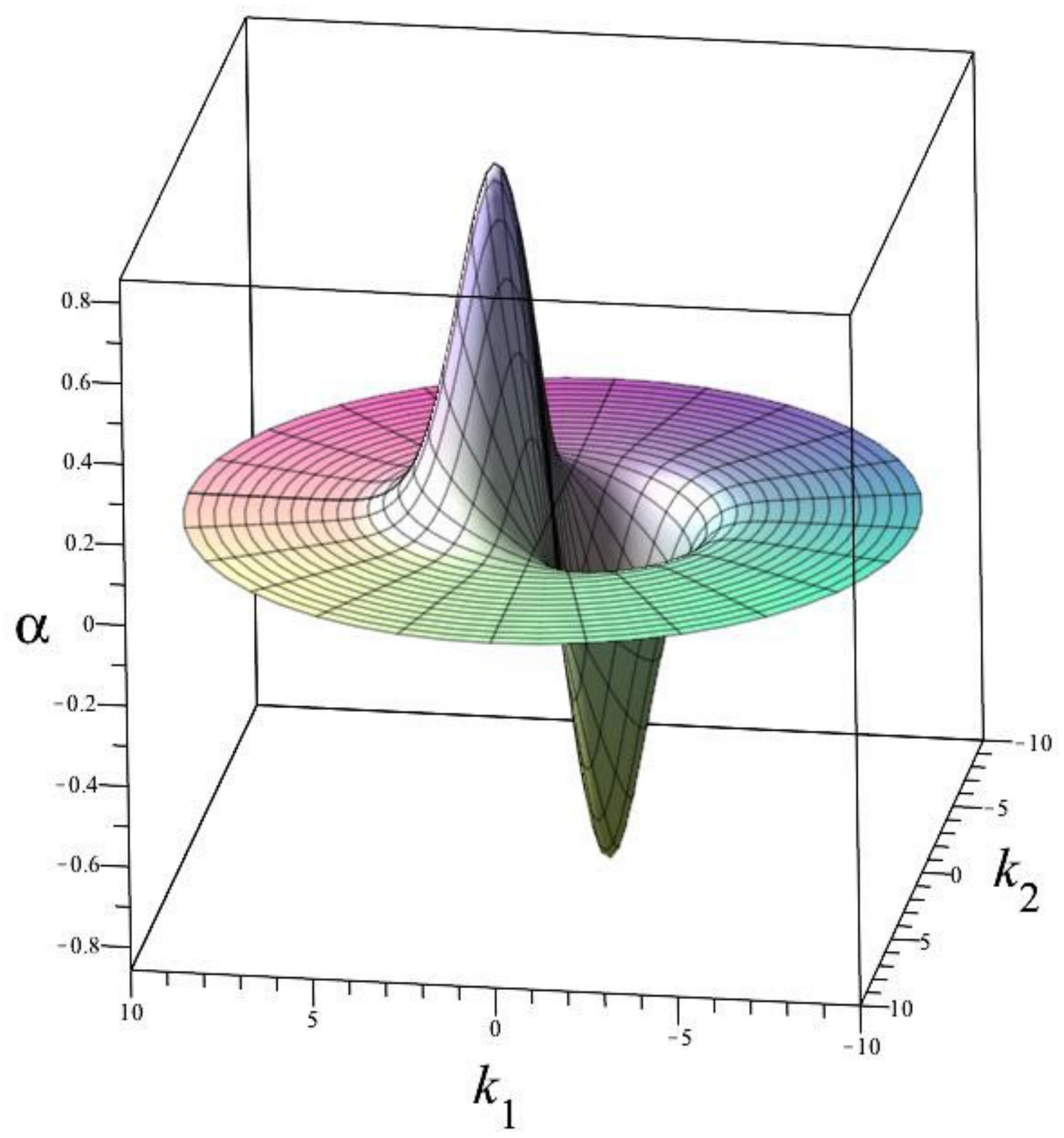}
\caption{\label{Fig1} Shape of $i \hat\alpha$, for N=1 and angular dependence of
cosine type. The deformation parameter $\theta$ is chosen to be
equal to one.}
\end{figure}
\item {N=2.} The corresponding amplitude is given by (see Eq.~\eqref{QN})  
\begin{equation}
Q_2 = \frac{1}{27\theta^2}.
\end{equation}
There are four linearly independent solutions \newline
\begin{equation*}
m=12, \quad n=1 \quad \mbox{angular dependence = cosine}
\end{equation*}
\begin{equation*}
m=12, \quad n=1 \quad \mbox{angular dependence = sine}\;\;\,
\end{equation*}
\begin{equation*}
m=0, \quad n=15 \quad \mbox{angular dependence = cosine}
\end{equation*}
\begin{equation*}
m=0, \quad n=15 \quad \mbox{angular dependence = sine}\;\;\,
\end{equation*}
In Fig. \ref{Fig2} we plot the corresponding potential with angular
dependence of cosine type.  
\begin{figure}[htb]
\centering
\includegraphics[scale = 0.33]{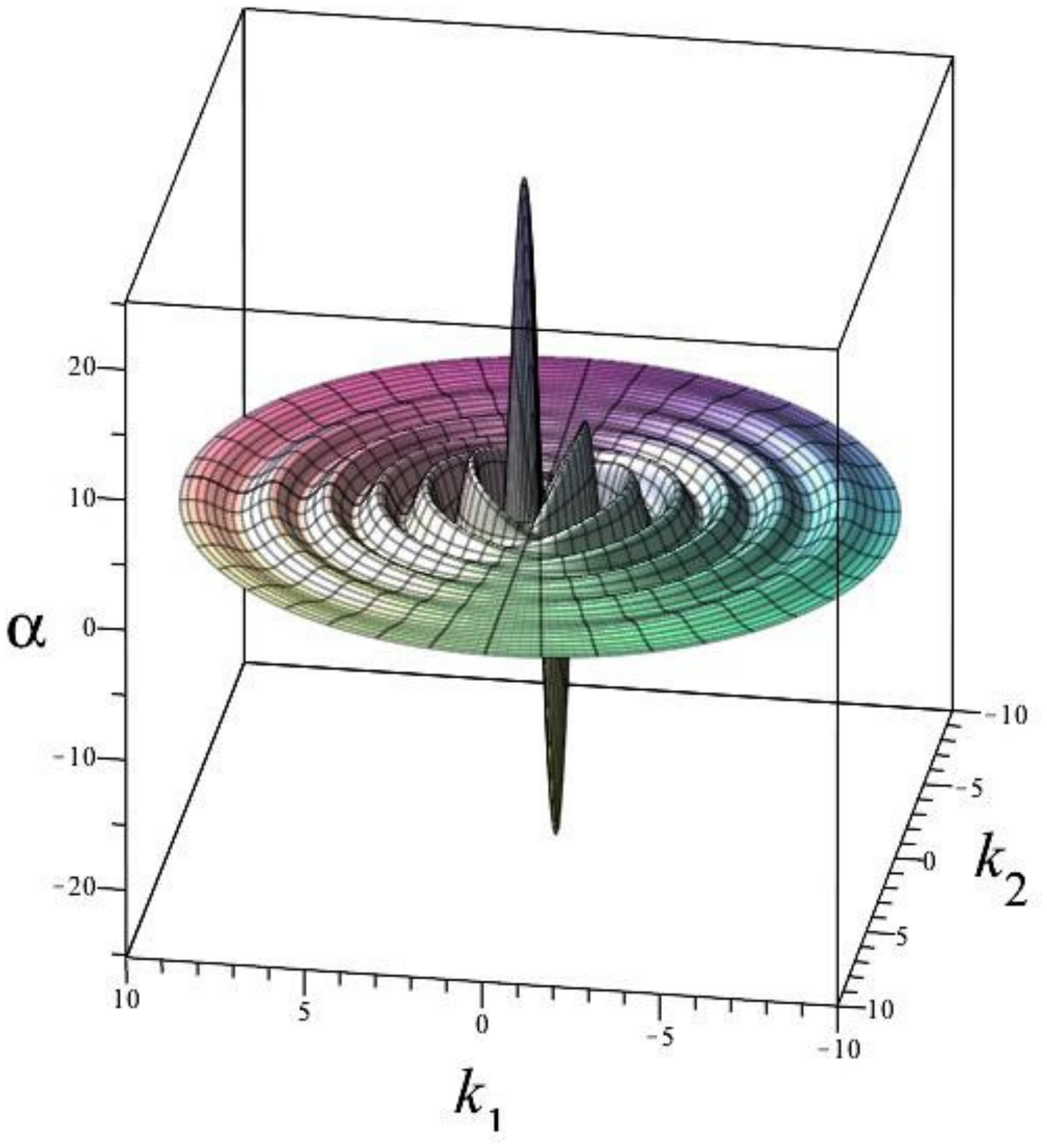}
\hfill
 \includegraphics[scale =0.33]{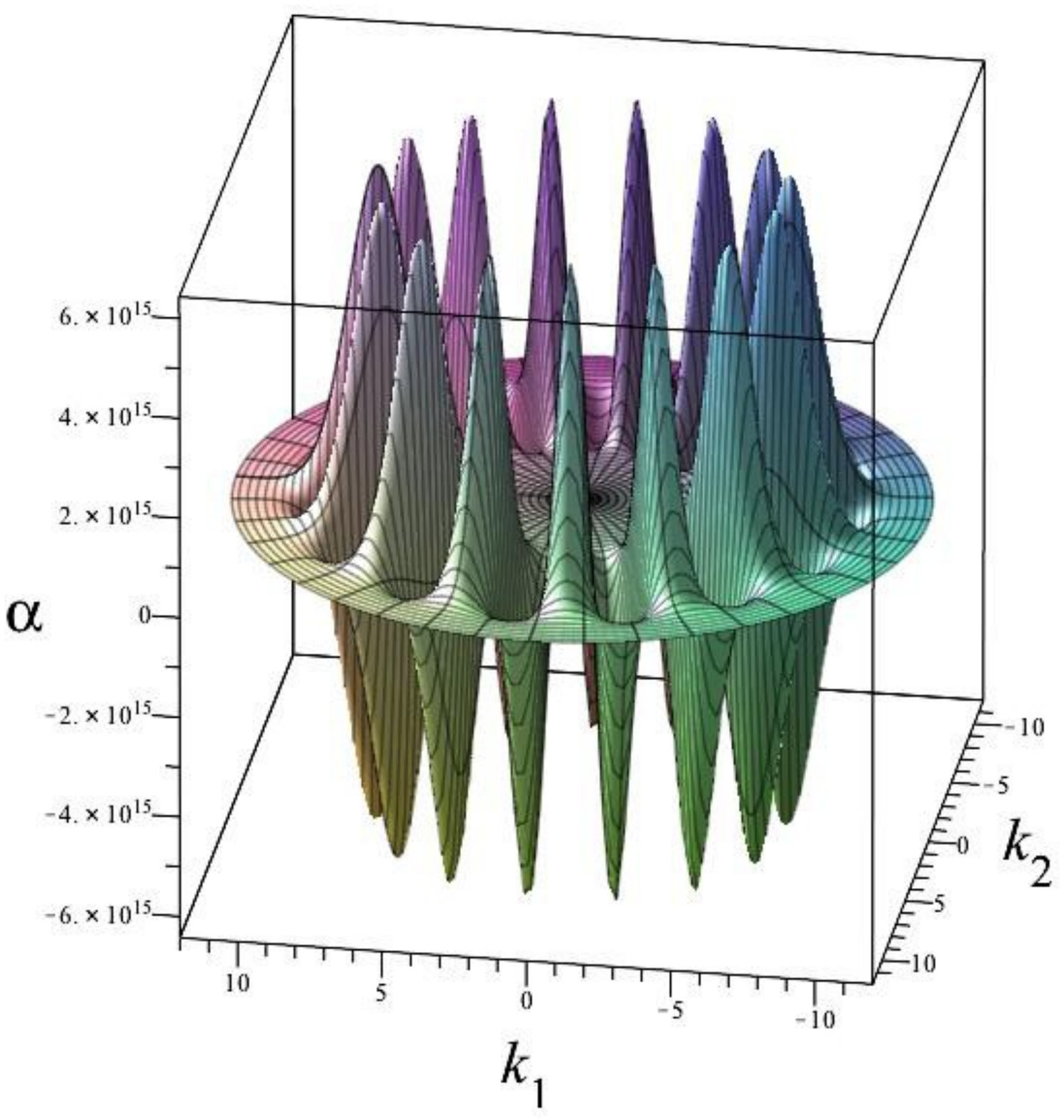}
\caption{\label{Fig2} Shape of $i \hat\alpha$, for N=2. The first plot represents
the case $m=12$, while the second one corresponds to $m=0$  of cosine type.
The deformation parameter $\theta$ is chosen to be equal to one.}
\end{figure}
\end{itemize}

\subsection{Multidimensional generalization}\label{multi}

Let us consider now the generalization to arbitrary dimensions of the
solutions discussed above.

Although the potential given by Eq.~\eqref{ntsp} satisfies the Landau
gauge fixing condition in arbitrary even dimensions, it does not lead to a
simple partial differential equation where one can easily separate
variables. 
However, if one considers the space ${\mathbb{R}}^d$ as a direct product of $%
d/2$ orthogonal planes, the tensor  $\epsilon^{\mu\nu}$ is invariant under
rotation in each plane. This observation suggests to  define the potential
by reproducing the two dimensional structure in each plane:  
\begin{equation}
\begin{cases}
A_1(x) = Q_{I} x_2 (x_1^2 +x_2^2) \\ 
A_2(x) = -Q_{I} x_1 (x_1^2 +x_2^2) \\ 
A_3(x) = Q_{I I} x_3 (x_3^2 +x_4^2) \\ 
A_4(x) = -Q_{I I} x_4 (x_3^2 +x_4^2) \\ 
.....................................,%
\end{cases}
\label{multidim}
\end{equation}
where $Q_{I},Q_{II}, ...$ are in general different constants.

Taking the Fourier transform of Eq.~\eqref{multidim} and substituting the
result in the integral equation \eqref{iEQ}, after carrying out similar
computations to the ones that we did in the $d= 2$ case, we arrive at the
following partial differential equation 
\begin{equation}
\left[Q_{I}D_{I} +Q_{II}D_{II} + ...\right]\hat\alpha =%
(k_{I}^2 + k_{II}^2+...)\hat\alpha ,  \label{multeq}
\end{equation}
where we used the following notations 
\begin{eqnarray}
&& k_{I}^2 = k_1^2 + k_2^2,  \notag \\
&& k_{II}^2 = k_3^2 + k_4^2,  \notag \\
&&.......................
\end{eqnarray}
and 
\begin{eqnarray}
&& D_{I} = -\theta_1 \left(k_{I}^2 \left(\partial_1^2 + \partial_2^2\right) +
2k_1^2\partial_2^2 + 2k_2^2\partial_1^2 - 4k_1k_2\partial_{12}^2 - \frac{%
\theta_1^2}{4}k_{I}^4\right)  \notag \\
&& D_{I I} = -\theta_2\left(k_{I I}^2 \left(\partial_3^2 + \partial_4^2\right) +
2k_3^2\partial_4^2 + 2k_4^2\partial_3^2 - 4k_3k_4\partial_{34}^2 - \frac{%
\theta_2^2}{4}k_{I I}^4\right)  \notag \\
&& ..........................
\end{eqnarray}
with $\theta_i, \, i=1,...,d/2$ the different noncommutative parameters of each two-dimensional plane, as in Eq. \eqn{nctheta}. One may easily see that Eq.\eqref{multeq} is a sum of $d/2$ equations 
\begin{eqnarray}
&& Q_{I}D_{I} \hat \alpha =
\hat\alpha k_{I}^2 ,  \label{pdEQ1} \\
&& Q_{II}D_{II} \hat \alpha =  
\hat \alpha k_{I I}^2 ,  \label{pdEQ2} \\
&& .......................  \label{pdEQ}
\end{eqnarray}
where each equation is of the form \eqref{PDE}.

If functions $\hat\alpha_{I}(k_1,k_2), \hat\alpha_{II}(k_3,k_4), ...$, are
solutions of \emph{two dimensional} equations Eq.~\eqref{pdEQ1}, Eq.~%
\eqref{pdEQ2}, ..., then their product  
\begin{equation}
\hat\alpha = \hat\alpha_{I}(k_1,k_2)\cdot \hat\alpha_{II}(k_3,k_4)\cdot ...
\label{alphamulti}
\end{equation}
solves each of the equations Eq.~\eqref{pdEQ1}, Eq.~\eqref{pdEQ2}, ... as
well as their sum Eq.~\eqref{multeq}.  But we have already constructed
solutions for the two dimensional case belonging to the Schwarz space $%
\mathcal{S}({\mathbb{R}}^2)$ hence the product Eq.~\eqref{alphamulti}
automatically belongs to the Schwarz space $\mathcal{S}({\mathbb{R}}^d)$.

\section{Discussion and outlook}

In the present paper we have shown that an infinite number of Gribov copies
 exists in noncommutative QED, and this is a genuine noncommutative
effect. As already recalled in the introduction and well known in the literature on noncommutative gauge theory, NCQED behaves like a non-Abelian gauge theory because of the non-trivial deformation of the covariant derivative Eq. \eqn{cd}. Therefore the existence of a nontrivial equation for copies was expected and not surprising. The main result of the paper is that such equation has solutions, which we compute,  and they are an infinite number.  The intrinsic interest of this observation is that it shows how
noncommutative geometry can give rise to a global obstruction preventing a
proper gauge-fixing already  in Abelian gauge theories. Consequently, the
intriguing possibility which is naturally suggested by the present analysis
is to extend to the case of NCQED the Gribov-Zwanziger restriction. Indeed,
the Gribov-Zwanziger restriction would yield to the following modification
of the NCQED propagator 
\begin{equation}
G^{\mathrm{G-Z}}(p)\sim \frac{p^{2}}{p^{4}+\gamma ^{4}}
\end{equation}%
with $\gamma $ depending on the noncommutative parameter $\theta $.

However, having proved that the noncommutativity of space-time induces
(infinitely many) Gribov copies also  in NCQED (as we did in the present
paper) is not enough to justify the Gribov-Zwanziger approach.

In non-Abelian gauge theory on flat topologically trivial space-times, the
Gribov-Zwanziger approach is based on the following fundamental results \cite%
{DZ89}:

\textbf{1)} The Gribov region is bounded in every direction (in the
functional space of transverse gauge potentials).

\textbf{2)} The Faddeev-Popov determinant changes sign at the Gribov horizon.

\textbf{3)} Every gauge orbit passes inside the Gribov horizon.

The importance of the last result lies in the fact that it justifies the
restriction of functional integration to the Gribov region (since what is
left outside the Gribov horizon is just a copy of something inside it and so
no relevant configuration is lost). 

In order to justify the Gribov-Zwanziger
restriction in NCQED we should generalize the analysis of \cite{DZ89} to the
noncommutative case. This is a highly non-trivial technical task since many
of the arguments used in these references to prove the properties \textbf{1)}%
, \textbf{2)} and especially \textbf{3)} make heavy use of the theory of local
elliptic PDEs while, in the noncommutative case, the Gribov copy equation
becomes non-local. We hope to come back in a future publication on these
important issues.

As a final remark, it is natural to wonder whether the present results can be extended to the Coulomb gauge as well. In the footnote \ref{Coulombgauge} we have described the equation of the copies in such a case.  Formally, some of the copies constructed here can also be used to construct copies in the Coulomb gauge (for instance, a copy in the Landau gauge in a two dimensional non-commutative space-time can be trivially promoted to a copy in the Coulomb gauge in 3+1 dimensions). However, the physical interpretation is rather subtle. Indeed, one of the fundamental properties of the Moyal star product is the `democracy' between all   space-time coordinates. Obviously, such democracy is not respected by the Coulomb gauge (one well known consequence being the existence of residual gauge transformations in this gauge). Thus, in the non-commutative setting, not only one should declare the Euclidean time as `special' but  its non-commutative partner  as well and this would make the analysis much more complicated than in the Landau case. We hope to come back on this interesting issue in a future investigation.

\subsection*{Acknowledgements}

F.C. acknowledges financial support from the Fondecyt grant no.1120352. The
Centro de Estudios Cient\'{\i}ficos (CECs) is funded by the Chilean
Government through the Centers of Excellence Base Financing Program of
Conicyt. M. K. and P.V.
acknowledge partial support from UniNA and Compagnia di San Paolo in the framework
of the program STAR 2013.  
M.K acknowledges partial support from FAPESP process 2015/05120-0.
L. R. acknowledges financial support by the program PRIN 2012 of the Italian Education and Research  Ministry (MIUR), project No.  
2012CPPYP7.


\begin{thebibliography}{99}
\bibitem{DFR03} S.~Doplicher, K.~Fredenhagen and J.~E.~Roberts,
  {\it The Quantum structure of space-time at the Planck scale and quantum fields, 
  Commun.\ Math.\ Phys.}  {\bf 172}, 187 (1995)
  [hep-th/0303037].\\
S.~Doplicher,
  {\it Space-time and fields: A Quantum texture, 
  AIP Conf.\ Proc.}  {\bf 589}, 204 (2001)
  [hep-th/0105251].


\bibitem{Br12} 
  M.~Bronstein,
   {\it  Quantum theory of weak gravitational fields,
  Gen.\ Rel.\ Grav. }  {\bf 44}, 267 (2012).

\bibitem{stringsNC1}  N.~Seiberg and E.~Witten,  {\it  String theory and
noncommutative geometry, JHEP}  \textbf{9909}, 032 (1999) 
[hep-th/9908142].  


\bibitem{As12} 
  A.~Ashtekar,
  {\it  Introduction to Loop Quantum Gravity,
  PoS QGQGS}  {\bf 2011}, 001 (2011)
  [arXiv:1201.4598 [gr-qc]].
\bibitem{Ro11}
  C.~Rovelli,
  {\it  Zakopane lectures on loop gravity,
  PoS QGQGS}  {\bf 2011}, 003 (2011)
  [arXiv:1102.3660 [gr-qc]].
  
  
  \bibitem{BO10} 
  A.~Baratin and D.~Oriti,
   {\it  Group field theory with non-commutative metric variables,
  Phys.\ Rev.\ Lett.}  {\bf 105}, 221302 (2010)
  [arXiv:1002.4723 [hep-th]].
  


\bibitem{DeW03} B. S. DeWitt, \textit{Global approach to quantum field
theory,} Vol. 1 and 2, Oxford University Press (2003).

\bibitem{Gri78} V.~N.~Gribov, {\it Quantization of Nonabelian Gauge Theories,
Nucl.\ Phys.} \textbf{ B 139} (1978) 1.

\bibitem{singer} I. M. Singer, {\it  Some remarks on the Gribov ambiguity,}
\textit{Comm. Math. Phys}. \textbf{60} (1978), 7.

\bibitem{lechtenfeld} P.~M.~Lavrov and O.~Lechtenfeld, {\it Gribov horizon
beyond the Landau gauge,  Phys.\ Lett.}  \textbf{ B 725}, 386 (2013) 
[arXiv:1305.2931 [hep-th]]. 

\bibitem{Zw82} D. Zwanziger, {\it Nonperturbative Modification
of the Faddeev-Popov Formula and Banishment of the Naive
Vacuum,\ Nucl.\ Phys.}  \textbf{ B 209} (1982) 336;\newline
{\it Action from the Gribov horizon, } \textit{Nucl. Phys}. \textbf{B 321}, (1989)
591;  {\it  Local and renormalizable action from the Gribov horizon,} \textit{Nucl.
Phys}. \textbf{B 323}, (1989) 513; {\it Renormalizability of the critical limit
of lattice gauge theory by BRS invariance, } \textit{Nucl. Phys.} \textbf{B 399%
}, (1993) 477.

\bibitem{DZ89} G. F. Dell'Antonio, D. Zwanziger, {\it  Ellipsoidal bound on the
Gribov horizon contradicts the perturbative renormalization group,  Nucl.
Phys.}  \textbf{B 326}, (1989) 333; {\it  Every gauge orbit passes inside the
Gribov horizon} \textit{Comm. Math. Phys.} \textbf{138}, 291-299 (1991).

\bibitem{Zwa96} D. Zwanziger, {\it  Renormalization in the Coulomb gauge and
order parameter for confinement in QCD,}  \textit{Nucl. Phys.} \textbf{B 518}
(1998) 237; {\it  No Confinement without Coulomb Confinement}  \textit{Phys. Rev.
Lett}. \textbf{90} (2003) 102001.

\bibitem{Va92} P. van Baal, {\it  More (thoughts on) Gribov copies, Nucl. Phys.}
\textbf{B 369}, (1992) 259.

\bibitem{MaggS} M. Maggiore, M. Schaden, {\it Landau gauge within the Gribov
horizon,}  \textit{Phys. Rev. }\textbf{D 50} (1994) 6616.

\bibitem{Gracey} J. A. Gracey, {\it One loop gluon form factor and freezing of
$\alpha_s$ in the Gribov-Zwanziger QCD Lagrangian,} \textbf{JHEP} 0605 (2006)
052.

\bibitem{SoVar} D. Dudal, S. P. Sorella, N. Vandersickel, H. Verschelde,
{\it New features of the gluon and ghost propagator in the infrared region from
the Gribov-Zwanziger approach,}  \textit{Phys. Rev. }\textbf{D 77} (2008)
071501.

\bibitem{SoVar2} D. Dudal, J. A. Gracey, S. P. Sorella, N. Vandersickel, H.
Verschelde, {\it  Refinement of the Gribov-Zwanziger approach in the Landau
gauge: Infrared propagators in harmony with the lattice results,} \textit{%
Phys. Rev. }\textbf{D 78} (2008) 065047.

\bibitem{SoVar3} D. Dudal, S. P. Sorella and N. Vandersickel, {\it  Dynamical
origin of the refinement of the Gribov-Zwanziger theory,} \textit{Phys. Rev}. 
\textbf{D 84}, 065039 (2011).

\bibitem{DOV} D. Dudal, O. Oliveira and N. Vandersickel, {\it  Indirect lattice
evidence for the refined Gribov-Zwanziger formalism and the gluon condensate 
$\langle A^2\rangle$ in the Landau gauge,}  \textit{Phys. Rev}. \textbf{D 81}
(2010) 074505.

\bibitem{soreprl} D. Dudal, M. S. Guimaraes, S. P. Sorella, {\it Glueball Masses
from an Infrared Moment Problem},  \textit{Phys. Rev. Lett.} \textbf{106},
062003 (2011).

\bibitem{CaRo} F. Canfora, L. Rosa,{\it Casimir energy in the
Gribov-Zwanziger approach to QCD,  Phys.\ Rev.}  \textbf{D 88%
}, 045025 (2013) [arXiv:1308.1582 [hep-th]]. 


\bibitem{CGO} F.~Canfora, A.~Giacomini and J.~Oliva, {\it  Gravitationally
induced zero modes of the Faddeev-Popov operator in the Coulomb gauge for
Abelian gauge theories, Phys.\ Rev.}  \textbf{D 82}, 045014 (2010).

\bibitem{ACGO} A. Anabalon, F.~Canfora, A.~Giacomini and J.~Oliva, {\it Gribov
ambiguity in asymptotically AdS three-dimensional gravity,}  \textit{Phys. Rev}%
. \textbf{D 83}, 064023 (2011).

\bibitem{CGO2} F. Canfora, A. Giacomini and J. Oliva, {\it Gribov pendulum in
the Coulomb gauge on curved spaces,} \textit{Phys. Rev}. \textbf{D 84},
105019 (2011).

\bibitem{giampi} M. de Cesare, G. Esposito, H. Ghorbani, {\it Size of the Gribov
region in curved spacetime,} \textit{Phys. Rev}. \textbf{D 88}, 087701 (2013).









\bibitem{heynyey} F. S. Henyey, {\it Gribov ambiguity without topological charge,}  \textit{Phys. Rev.} \textbf{D 20}, 1460
(1979).

\bibitem{wallet} J.-C. Wallet, {\it Derivations of the Moyal algebra and
noncommutative gauge theories, SIGMA}{\bf  5} (2009) 013 [arXiv:0811.3850]. 
\newline
E. Cagnache, T. Masson and J.-C. Wallet, {\it Noncommutative Yang-Mills-Higgs
actions from derivation-based differential calculus, J. Noncomm. Geom.}  {\bf 5}
(2011) 39 [arXiv:0804.3061]. 

\bibitem{MVW13} P.~Martinetti, P.~Vitale and J.~C.~Wallet,  {\it Noncommutative
gauge theories on $\mathbb{R}^2_\theta$ as matrix models,  JHEP} \textbf{%
1309}, 051 (2013)  [arXiv:1303.7185 [hep-th]].  

\bibitem{A001353} 
\url
{https://oeis.org/A001353}

\bibitem{Bat} A. Erdelyi, {\it Higher Transcendental Functions,}  Bateman
Manuscript Project, California Institute of Technology, Vol. 1, ISBN
0-486-44614-X


\end{thebibliography}
\end{document}